\documentclass[letterpaper,floatfix,showpacs,aps,prb,twocolumn,10pt,superscriptaddress,showpacs]{revtex4}

\usepackage{latexsym,amsmath,amssymb,amsfonts,mathbbol,graphicx}
\usepackage{dcolumn}
\usepackage{bm}
\usepackage[T1]{fontenc}
\usepackage[latin1]{inputenc}
\usepackage[usenames,dvips]{color}

\begin{document}

\title{Anomalous non-ergodic scaling in adiabatic multicritical quantum
quenches}

\author{Shusa Deng}
\affiliation{\mbox{Department of Physics and Astronomy, Dartmouth
College, 6127 Wilder Laboratory, Hanover, New Hampshire 03755,
USA}}

\author{Gerardo Ortiz}
\affiliation{Department of Physics, Indiana University,
Bloomington, Indiana 47405, USA}

\author{Lorenza Viola}
\affiliation{\mbox{Department of Physics and Astronomy, Dartmouth
College, 6127 Wilder Laboratory, Hanover, New Hampshire 03755,
USA}}

\date{\today}

\begin{abstract}
We investigate non-equilibrium dynamical scaling in adiabatic
quench processes across quantum multicritical points.  Our
analysis shows that the resulting power-law scaling {\it depends
sensitively on the control path}, and that anomalous critical
exponents may emerge depending on the universality class.  We
argue that the observed anomalous behavior originates in the fact
that the dynamical excitation process takes place asymmetrically
with respect to the static multicritical point, and that
non-critical energy modes may play a dominant role.  As a
consequence, dynamical scaling requires introducing new non-static
exponents.
\end{abstract}


\maketitle


Establishing dynamical scaling relations in many-body systems
adiabatically driven out-of-equilibrium across a quantum phase
transition has important implications for both condensed-matter
physics \cite{Sachdev}, and adiabatic quantum computation
\cite{Adiabqc}.  A paradigmatic scenario is offered by the
so-called {\em Kibble-Zurek scaling} (KZS) \cite{Zurek,Anatoli},
whereby an homogeneous $d$-dimensional system is linearly driven
with a constant speed $1/\tau$ across an isolated quantum critical
point (QCP) described by equilibrium critical exponents $\nu$ and
$z$.  Under the assumption that, in the thermodynamic limit, the
system loses adiabaticity throughout an ``impulse region''
$[t_c-\hat{t}, t_c+\hat{t}]$ which is centered around the QCP and
has a characteristic width 2$\hat{t}$, excitations are generated
in the final state with a density $n_{\text{ex}}(t_f) \sim
\tau^{-d\nu/(\nu z+1)}$.  While the KZS and its non-linear
generalizations have been verified in several exactly solvable
models \cite{KZSMix}, departures from the KZ prediction may occur
for more complex quench processes -- involving either isolated
QCPs in disordered \cite{Disorder} and infinitely-coordinated
systems \cite{Tommaso}, or non-isolated QCPs (that is, quantum
critical regions) \cite{Pellegrini,deng1,chowdhury}. Evidence of
non-KZS, however, has also been reported in the apparently simple
situation of a quench across a single quantum {\em multicritical
point} (MCP) in clean spin chains \cite{Victor2,Sen2}.

In this work, we show how multicritical quantum quenches
dramatically exemplify the dependence of non-equilibrium scaling
upon the control path anticipated in Ref. \onlinecite{deng1}, and
demonstrate that anomalous ``non-ergodic'' scaling may emerge in
the thermodynamic limit.  While a non-KZS $n_{\text{ex}}(t_f) \sim
\tau^{-1/6}$ was previously reported \cite{Victor2} and an
explanation given in terms of an ``effective dynamical critical
exponent'' $z_2=3$, the meaning of such exponent relied on the
applicability of a Landau-Zener (LZ) treatment, preventing general
insight to be gained.  We argue that the failure of KZS is
physically rooted in the shift of the center of the impulse region
relative to the static picture, and that $z_2$ is determined by
the scaling of a path-dependent {\em minimum gap} which need not
coincide with the critical gap.  Furthermore, such a dynamical
shift may also cause the contribution from {\em intermediate
non-critical energy states} to dominate the scaling of the
excitation density, via an ``effective dimensionality exponent''
$d_2 \ne 0$. We show that the latter leads to the emergence of new
scaling behavior $n_{\text{ex}}(t_f) \sim \tau^{-3/4}$.  A unified
understanding is obtained by extending the adiabatic
renormalization (AR) approach of Ref. \onlinecite{deng1}.


{\em Model Hamiltonian.$-$} We focus on the alternating spin-$1/2$
XY chain described by the Hamiltonian \cite{deng1,deng2}:
\begin{eqnarray}
H = - \hspace*{-0.8mm}\sum_{i=1}^N (\gamma_+ \sigma_x^i
\sigma_x^{i+1} \hspace*{-0.8mm} + \gamma_-  \sigma_y^i
\sigma_y^{i+1} - \hspace*{-0.7mm} h_i \sigma_z^i ), \label{Ham}
\end{eqnarray}
where $\gamma_\pm=({1\pm\gamma})/2$, $h_i=h -(-)^i \delta $, and
periodic boundary conditions are assumed.  Here, $h, \delta \in
\mathbb{R}$ are the uniform and alternating magnetic field
strength, respectively, whereas $\gamma \in \mathbb{R}$ is the
anisotropy (lifting the restriction $\gamma \in [0,1]$ is
essential for the present analysis).  An exact solution for the
energy spectrum of $H$ may be obtained through the steps outlined
in Ref. \onlinecite{deng2}. The problem maps into a collection of
non-interacting quasi-particle labelled by momentum modes $k \in
K_+ =\{\pi/N,3\pi/N,\ldots, \pi/2-\pi/N \}$, whose excitation gap
is given by $ \Delta_k (\gamma, h, \delta) = 4\,[h^2+\delta^2 +
\cos^2 k+\gamma^2 \sin^2 k -
2\sqrt{h^2\cos^2k+\delta^2(h^2+\gamma^2\sin^2k)}]^{1/2}$.  The
quantum phase boundaries are determined by the equations
\cite{deng1,deng2} $h^2 =\delta^2+1$, $\delta^2 = h^2+\gamma^2$.
Thus, the critical lines on the $\gamma=0$ plane consist entirely
of MCPs.

\begin{figure}[h]
\centering
\includegraphics[width=8cm]{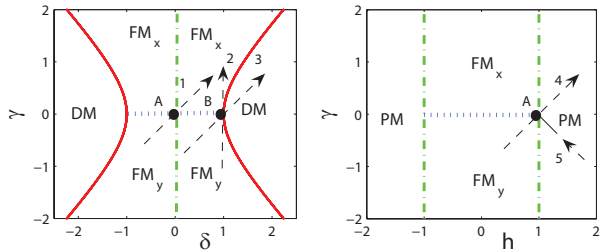}
\caption{Phase diagram of $H$ in Eq.~(\ref{Ham}) when $h=1$ (left)
and $\delta=0$ (right). The dashed-dotted (green) line separates
the ferromagnetic (FM) and paramagnetic (PM) phases, the solid
(red) lines separate dimer (DM) and FM, whereas the dotted (green)
line is the superfluid phase (SF). The arrows indicate the control
paths we choose to approach the MCPs {\tt A} and {\tt B}. }
\label{multi}
\end{figure}

{\em Quench dynamics: Exact results.$-$} We assume that the system
is initially in the ground state, and that (in the simplest case)
a slow quench across a MCP is implemented upon changing a single
control parameter according to $\delta
\lambda(t)=\lambda(t)-\lambda_c = |(t-t_c)/\tau|^\alpha
\mbox{sign}(t-t_c)$ over a time interval $t \in [t_0, t_f]$, where
$\alpha=1$ corresponds to a linear driving, and $\lambda_c$ is the
critical value.  Thus, the time-dependent Hamiltonian $H(t)$ may
be written as: $H(t)=H_c+\delta \lambda(t)H_1$, where $H_c$ is
quantum-multicritical at time $t_c$ in the thermodynamic limit,
and $H_1$ is the contribution that couples to the external control
(a similar parametrization is possible for quenches involving
multiple parameters). Without loss of generality, we may let
$t_c=0$. In what follows, we shall focus on two representative
MCPs, {\tt A} and {\tt B} as marked in Fig.~\ref{multi}, each
approached through two different paths (path $5$ will be
introduced later) with the following properties:
\begin{table}[ht]
\centering
\begin{tabular}{c|cc|l}
    {\tt Path }  &  $\nu$ & $z$ & {\tt \hspace{12mm}Quenching scheme} \\\hline
1 &  1 & \ 2 & $\gamma(t)=\delta(t)=|t/\tau|^\alpha \mbox{sign}(t);\; h=1$   \\
2 &  1 & \ 2 & $\gamma(t)=|t/\tau|^\alpha \mbox{sign}(t);\; h=1,\delta=1$  \\
3 &  1/2 & \ 2 & $\gamma(t)=\delta(t)-1 = |t/\tau|^\alpha
\mbox{sign}(t); \;
h=1$ \\
4 & 1/2 & \ 2 & $\gamma(t)=h(t)-1=|t/\tau|^\alpha \mbox{sign}(t);\; \delta=0$  \\
\hline
\end{tabular}
\end{table}

\begin{figure}[t]
\centering
\includegraphics[width=8cm]{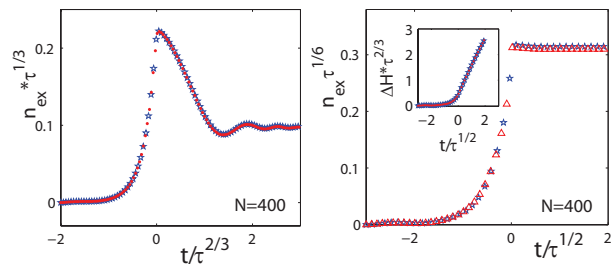}
\caption{Exact scaling of the excitation density throughout a
linear quench along path $2$ (left) and path $3$ (right). Right
inset: Scaling of the residual energy along path $3$.}
\label{mul24}
\end{figure}

In order to quantify the amount of excitation at a generic instant
$t$, we numerically integrate the time-dependent Schr\"odinger
equation for $H(t)$ and monitor two standard ``non-adiabaticity''
indicators \cite{deng2,deng1,Tommaso}: the excitation density,
$n_{\text{ex}}$, and the residual energy, $\Delta H$.
For a linear quench along either path $1$ or $2$ (left panel of
Fig.~\ref{mul24}), we find that $n_{\text{ex}}(t)\sim
\tau^{-\nu/(\nu z+1)}= \tau^{-1/3}$ and $\Delta H(t) \sim
\tau^{-\nu (1+z)/(\nu z+1)}=\tau^{-1}$,
which is consistent with KZS \cite{Zurek} and our conclusion in
Ref. \onlinecite{deng1}. For paths $3$ and $4$, however (right
panel of Fig.~\ref{mul24}), we find that $n_{\text{ex}}(t) \sim
\tau^{-1/6}$ and $\Delta H (t) \sim \tau^{-2/3}$, which is non-KZS
(in Ref. \onlinecite{Victor2}, the $\tau^{-1/6}$ scaling was
pointed out for an equivalent quench scheme across MCP {\tt A}).
Similar anomalous exponents are found for non-linear quenches
along paths 3 or 4, {\em e.g.}, $n_{\text{ex}}(t) \sim
\tau^{-2/9}$ for $\alpha=2$.

\begin{figure}[ht]
\centering
\includegraphics[width=8cm]{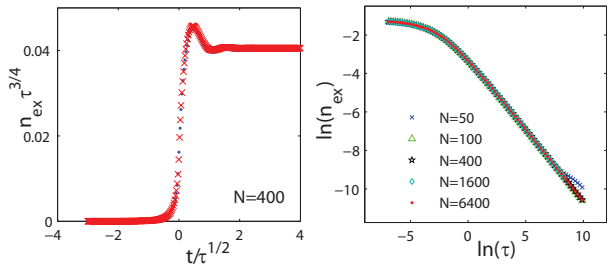}
\vspace{2mm} \caption{Left: Exact scaling of the excitation
density throughout a linear quench along path $5$. Right: Scaling
of the final excitation density in path $5$ for different system
size: $n_{\text{ex}}(t_f)$ is the same to a numerical accuracy of
$10^{-6}$, up to $\tau < 2\times10^5$. A linear fit yields $-0.747
\pm 0.001$ over the range $200 < \tau < 2000$.} \label{path5}
\end{figure}

The above results clearly show that, for quenches across a MCP,
whether KZS is obeyed depends sensitively on which control path is
chosen.  A closer inspection reveals the following important
differences: (i) Paths $1,2$ start and end in essentially the same
phase, correspondingly the excitation spectrum is invariant under
a transformation $\lambda \mapsto -\lambda$ of the control
parameters. Paths $3,4$ do not exhibit this symmetry; (ii) Along
paths $3,4$, the MCPs {\tt A} and {\tt B} belong to the Lifshitz
universality class ($\nu=1/2$), although all paths share $z=2$.
It is then natural to ask which of these differences may play a
role in determining the anomalous dynamical scaling behavior.  To
answer this question, we introduce another path across MCP {\tt A}
(path $5$), $h(t)=1+|\gamma(t)|=1+|t/\tau|$, which starts and ends
in the PM phase but, in each of the two segments, crosses the MCP
{\tt A} with Lifshitz exponents.  Surprisingly, the observed
scaling is $n_{\text{ex}}(t) \sim \tau^{-3/4}$ (left panel of
Fig.~\ref{path5}), which is neither KZS nor $-1/6$.  An identical
$-3/4$ scaling holds for a ``V-shaped'' path across MCP {\tt B},
that starts and ends in the DM phase.  As finite-size analysis
reveals, all the observed anomalous scalings are practically
independent upon system size over a wide range of quench rates
(see {\em e.g.} right panel of Fig.~\ref{path5}), establishing
them as truly thermodynamic in nature \cite{remark2}.


{\em Landau-Zener analysis.$-$} We begin to seek an understanding
from limiting cases where an exact solution for
$n_{\text{ex}}(t_f)$ may be obtained based on the LZ picture
\cite{Zener}. This is possible provided that $\alpha=1$ and the
Hamiltonian can be decoupled into effective two-level systems.
Among the above-mentioned paths, only paths $4$ and $5$ (for which
$\delta =0$) can be exactly mapped to a LZ problem, thanks to the
possibility of rewriting $H$ in Eq.~(\ref{Ham}) as $H=\sum_{k}
\hat{H}_k=\sum_{k} {B}_k^{\dag} {H}_k {B}_k$, where
${B}_k^\dag=(c_{-k},c_{k}^\dag)$ and
\begin{eqnarray}
{H}_{k}=\hspace{-1mm}\left (\hspace{-1mm}
\begin{array}{cc} H_{k,11} & H_{k,12} \\
H_{k,12}^{\ast} & -H_{k,11}
\end{array} \hspace{-1mm}\right)
\hspace{-1mm}=2\hspace{-0.5mm} \left( \hspace{-1mm}
\begin{array}{cc}
-h +\cos k & \gamma\sin k \\ \gamma\sin k & h-\cos k
\end{array} \hspace{-1mm}\right).
\label{twobytwo}
\end{eqnarray}
A rotation $R_k(q_k)$, $q_k \in [-\pi/2,\pi/2)$, renders the
off-diagonal terms in Eq.~(\ref{twobytwo}) independent upon
$\gamma$ (hence time), allowing use of the LZ formula. Consider
path $4$ first. By choosing $\tan 2q_k =-\sin k$, the transformed
Hamiltonian matrix elements become $H'_{k,11}=-2(1-\cos k)\cos
2q_k - 2t/\tau(\cos 2q_k -\sin k \sin 2q_k)$, and
$H'_{k,12}=2(1-\cos k) \sin 2q_k$.  If the critical mode $k_c$ is
defined by requiring $\Delta_{k_c} =0$ in the thermodynamic limit,
we have $k_c=0$ for the MCP ${\tt A}$.  We may then let $\tan 2q_k
\approx \sin 2q_k$, and the appropriate $q_k \approx -k/2$.  From
the LZ formula, the asymptotic ($t_f \rightarrow \infty$)
excitation probability reads
\begin{eqnarray*}
p_k\hspace*{-0.5mm}=\hspace*{-0.4mm} e^{-2\pi \tau (1-\cos k)^2
\sin^2 {2q_k}/(\cos 2q_k-\sin k \sin 2q_k)}\hspace*{-0.5mm}\approx
\hspace*{-0.5mm} e^{-\pi \tau k^6/2},
\end{eqnarray*}
where the approximation follows from a Taylor expansion around
$k_c$. Integrating over all modes yields $n_{\text{ex}}(t_{f})
\sim \tau^{-1/6}$, which is consistent with our exact numerical
result. Therefore, mathematically, the $\tau^{-1/6}$ scaling
follows from the fact that the exponent in $p_k$ scales as
$k^6=k^{2z_2}$, with $z_2=3$. In turn, this originates from the
scaling of the off-diagonal terms $H'_{k,12} \sim k^{z_2}$.
Physically, as we shall later see by invoking AR, $H'_{k,12}$ may
be interpreted as the minimum gap for mode $k$ along path $4$.

To unveil the $\tau^{-3/4}$ scaling, it is necessary to use the
{\em exact finite-time} LZ solution. For simplicity, we restrict
to half of path $5$, by quenching the system from the PM phase up
to the MCP {\tt A}.  This has the benefit of avoiding the
non-analytic time-dependence of the control parameters that path
$5$ exhibits at {\tt A}, while leaving the $\tau^{-3/4}$ scaling
unchanged thanks to the symmetry of the excitation spectrum.
Starting from Vitanov's expression [Eq.  (7) in Ref.
\onlinecite{Vitanov}], the excitation probability $p_k(t_f)$ can
be computed via the parabolic cylinder function $D_v(z)$,
\begin{eqnarray*}
p_k(t_f) &\hspace*{-0.5mm}= \hspace*{-0.5mm}& e^{-\pi
\omega^2/4}\Big{|}D_{i\omega^2/2}(T_f \sqrt{2}e^{i 3\pi/4})\cos
\theta(T_f) \\ &\hspace*{-0.5mm}-\hspace*{-0.5mm}&
\frac{\omega}{\sqrt{2}}e^{-i \pi/4}D_{i\omega^2/2-1}(T_f
\sqrt{2}e^{i 3\pi/4})\sin \theta(T_f)) \Big{|}^2,
\end{eqnarray*}
where $\omega=(1-\cos k)\sin 2 {q_k} \sqrt{\tau} /\sqrt{\cos 2
{q_k} +\sin 2 {q_k} \sin k} \sim k^3\sqrt{\tau} $ is the rescaled
coupling strength, $T_f=-\omega/\sin k \sim -k^2\sqrt{\tau}$ is
the rescaled time, $\tan {2 q_k}=\sin k$, and
$\theta(T_f)={1}/{2}\arctan{(\omega/T_f)}+\pi/2$.
Since for our quench process $\omega \ll |T_f| \ll 1$ around
$k_c$, we may estimate $p_k(t_f)$ by Taylor-expanding $D_v(z)$
around $T_f=0$:
\begin{eqnarray}
p_k(t_f)&\hspace*{-0.5mm}\approx\hspace*{-0.5mm}& (1 -
e^{-\pi\omega^2/2})/2+\cos^2{\theta(T_f)}e^{-\pi\omega^2/2}
\nonumber
\\
&\hspace*{-0.5mm}-\hspace*{-0.5mm}&\sin{2\theta(T_f)}/2\sin{\chi_k}\sqrt{1
- e^{-\pi\omega^2}}, \label{appk}
\end{eqnarray}
where $\chi_k\approx\pi/4$ around $k_c$.  This approximation
breaks when $T_f \sim 1$, setting the scaling of the
highest-momentum contributing mode, that is, $k_{\text{max}} \sim
\tau^{-1/4}$. In Eq.~(\ref{appk}), the dominant term
$\cos^2{\theta(T_f)}e^{-\pi\omega^2/2} \sim \cos^2{\theta(T_f)}
\sim |\omega/T_f|^2 \sim k^2$ since $e^{-\pi\omega^2/2} \approx 1$
within $k_{\text{max}}$, which means $p_k(t_f) \sim k^2$.  Thus,
$n_{\text{ex}}(t_f) = \int_0^{k_{\text{max}}} p_k(t_f) \sim
k_{\text{max}}^3 \sim \tau^{-3/4}$, in agreement with our
numerical results.  Remarkably, the fact that $p_k (t_f) \sim
(k-k_c)^{d_2}$, $d_2=2$, indicates that $k_c$ is {\em not} excited
despite a static QCP being crossed, and also that the excitation
is dominated by intermediate energy states. In fact, at the MCP
{\tt A}, the modes around $k_c$ are still far from the impulse
region, since $|T_f| \gg \omega$, which sets the LZ transition
time scale \cite{Vitanov}.  This is in stark contrast with the
main assumption underlying KZS, where the center of the impulse
region is the static QCP, and excitations are dominated by modes
near $k_c$, as reflected in the typical scaling $p_k \sim
(k-k_c)^0$.  Therefore, the shift of the actual (dynamical)
impulse region relative to the static one is ultimately
responsible for the anomalous $\tau^{-3/4}$ scaling.

\begin{figure}[tb]
\centering
\includegraphics[width=8cm]{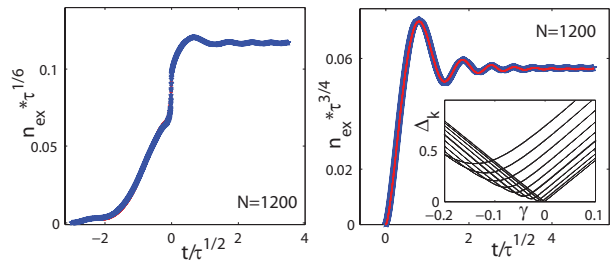}
\vspace{2mm} \caption{Scaling of the excitation density from
first-order AR for a linear quench along path $3$ (left) and
half-$5$ starting at MCP {\tt A} (right).  Right inset: Low-lying
single-mode excitation spectrum along path 4 for $N=100$.}
\label{tpm}
\end{figure}

{\em Perturbative scaling approach.$-$} Since the system becomes
gapless at a single MCP along all the paths under study,
first-order AR is a viable approach \cite{Anatoli,deng1}.  Let
$|\psi_m (t)\rangle$ be a basis of snapshot eigenstates of $H(t)$,
with snapshot eigenvalues $E_m(t)$, $m=0$ labelling the ground
state. The time-evolved state may be expanded as $|\psi(t)\rangle
= c_0^{(1)} (t) |\psi_{0} (t)\rangle + \sum_{m \ne 0} c_m^{(1)}(t)
|\psi_m (t)\rangle$, where the coefficients $c_m^{(1)}(t)$
determine the excitation amplitudes and are given by Eq.~(4) in
Ref. \onlinecite{deng1}. First-order AR calculations of
$n_{\text{ex}}(t)$ demonstrate that for linear quenches along
paths $1$ and $2$, $n_{\text{ex}} \sim \tau^{-1/3}$, whereas
$n_{\text{ex}} \sim \tau^{-1/6}$ along paths $3$ and $4$ (left
panel of Fig.~\ref{tpm}).
Since the non-analyticity at {\tt A} in path $5$ might cause
problems in AR, again we choose to study half of path $5$ (right
panel of Fig.~\ref{tpm}). All the AR results agree with the exact
simulation results, confirming that AR reproduces the correct
dynamical scaling across a {\em generic} isolated QCP.

Predicting the scaling exponent based on AR requires scaling
assumptions for the contributions entering $c_m^{(1)}(t)$ [{\em
i.e.}, $\Delta_m(t)=E_m(t)-E_0(t)$ and $\langle
\psi_m(t)|H_1|\psi_0(t)\rangle$] and the ability to change
discrete sums of all the contributing excited states into
integrals, for which the density of excited states $\rho(E)$ is
required. Since typically the AR prediction is consistent with
KZS, anomalous behavior must stem from anomalous scaling
assumptions of (one or more of) these ingredients. We first
examine the excitation spectrum along different paths. Notice that
since $H_1$ is a one-body perturbation, only single-mode
excitations are relevant, thus the index $m$ labeling many-body
excitations may be identified with a momentum mode.  Along paths
$1$ and $2$, it turns out that the minimum gap among all modes is
{\em always} located at $k_c$, whereas along paths $3$ and $4$,
the minimum gap is located at $k_c$ {\em only} at the MCP.  This
suggests that knowing the critical exponents of the MCP alone need
not suffice to determine the dynamical scaling due to the
existence of ``quasi-critical'' modes along paths $3$ and $4$. In
fact, mathematically, along path $4$,
$\partial{\Delta_k(\gamma,1+\gamma,0)}/\partial{\gamma}=0$ gives
the location of the minimum gap for each mode $k$ at
$\tilde{\gamma}=(\cos{k}-1)/(1+\sin{k}^2)$, which is largely
shifted into the FM phase (see inset in Fig.~\ref{tpm}). By
inserting this relation back into $\Delta_k$, the composite
function $\Delta_k(\tilde{\gamma}) \equiv \tilde{\Delta}_k \sim
(k-k_c)^3$. Following the same procedure also yields
$\tilde{\Delta}_k \sim (k-k_c)^3$ along path $3$, while
$\tilde{\Delta}_k$ has the same scaling as $\Delta_k$ at the MCP
along paths $1$ and $2$.  This motivates modifying the AR scaling
assumptions of Ref. \onlinecite{deng1} as follows: $E_m
(t)-E_0(t)= \delta\lambda(t)^{\nu z} f_m ({\Delta_m
(t_{\text{min}})}/{\delta\lambda(t)^{\nu z}})$, where $\Delta_m
(t_{\text{min}})$ is the minimum gap of mode $m$ attained at
$t_{\text{min}}$ along the path, and $f_m$ is a scaling function.

The above modification requires the scaling of $\rho(E)$ to be
modified by letting $\rho(E) \sim E^{d/z_2-1}$, where $z_2$ comes
from the dispersion relation of $\Delta_m (t_{\text{min}})$.  If
the minimum gap of any mode is below a certain energy along the
path, that mode should be counted into the contributing excited
states. Accordingly, we have $z_2=z=2$ along paths $1$, $2$,
half-$5$, and $z_2=3$ along paths $3$, $4$.  Back to the LZ
analysis, note that the off-diagonal term $H'_{12}(k)$ {\em is}
the minimum gap of mode $k$ along the path if there exists a time
at which the diagonal term $H'_{11}(k)=0$, as it happens for path
$4$.  For path $5$, however, the off-diagonal term never becomes
the minimum gap since the system never leaves the PM phase.
Therefore, the off-diagonal term in the LZ picture need not
suffice to determine the dynamical scaling, and the shift of the
location of the minimum gap for each mode from the static QCP is
at the root of the anomalous behavior we observe.  Lastly, we
consider the matrix elements of $H_1$.  Numerical simulations
suggest that $\langle \psi_m (t)|H_1| \psi_0
(t)\rangle=\delta\lambda(t)^{\nu z-1} g_m (\Delta_m
(t_{\text{min}}) /\delta\lambda(t)^{\nu z})$, where $g_m$ is a
scaling function, and $\Delta_m(t_{\text{min}})$ is the minimum
gap of mode $m$ along a path that {\em extends} the actual path to
$t_f \rightarrow \infty$ when the quench is stopped at the MCP,
and coincides with the actual path otherwise. Then along paths $1$
and $2$, $\Delta_m (t_{\text{min}}) \sim k^2$, whereas along paths
$3$, $4$, and half-$5$, $\Delta_m (t_{\text{min}}) \sim k^3$.
Together with the other scaling assumptions, and taking the linear
case $\alpha=1$ as an example, AR yields $|c_m^{(1)}| \sim k^0$,
$n_{\text{ex}} \sim \tau^{- (z/z_2) (\nu /(\nu z+1))}$ along paths
$1$ to $4$, and $|c_m^{(1)}| \sim k^1$, $n_{\text{ex}} \sim
\tau^{-3\nu /(\nu z+1)}$ along half-$5$ path, which completely
agrees with the numerical results.

Building on the above analysis, we argue on physical grounds that
the scaling of the excitation density for quenches across an {\em
arbitrary (standard or multicritical) isolated QCP} is determined
by three conditions: (i) From the condition of adiabaticity
breaking, the typical gap, $\hat{\Delta}$, scales as $\hat{\Delta}
\sim \tau^{-\alpha\nu z/(\alpha\nu z+1)}$; (ii) An accessible
excited state contributes to the excitation if and only if its
minimum gap along the path matches with this typical gap,
$\tilde{\Delta}_k \sim \hat{\Delta}$, with $\tilde{\Delta}_k \sim
(k-k_c)^{z_2}$;
(iii) The excitation probability, $p_k$, scales as $p_k \sim
(k-k_c)^{d_2}$, where $d_2$ can differ from $0$ if the center of
the impulse region is greatly shifted relative to the static
limit.
Then upon integrating up to energy $\hat{\Delta}$, and using
$p_E\sim p_{k(E)} \sim E^{d_2/z_2}$, yields \vspace*{-2mm}
\begin{equation}
n_{\text{ex}} \sim \hat{\Delta}^{(d+d_2)/z_2} \sim \tau^{-(d+d_2)
\alpha\nu z /[z_2 (\alpha\nu z+1)]}, \label{nexg}
\end{equation}
which is consistent with all the results found thus far.  While
KZS corresponds to $z_2=z$, $d_2=0$, situations where $z_2 \neq z$
and/or $d_2 \neq 0$ are {\em genuinely dynamical}: knowledge about
the path-dependent excitation process becomes crucial and
non-equilibrium exponents cannot be fully predicted from
equilibrium ones. Interestingly, in the model under examination
the Lifshitz universality class appears to be the only
universality class for which anomalous scaling occurs, among all
possible paths involving MCPs. Whether Lifshitz behavior may
constitute a {\em sufficient} condition for anomalous behavior
requires further investigation in other many-body systems.

It is a pleasure to thank Tommaso Caneva for insightful exchange.
S.D. gratefully acknowledges support from
a Hull Graduate Fellowship.


{}


\begin{thebibliography}{100}
\bibitem{remark2} Since $\nu z < z$, to leading order the scaling of
$\Delta_k$ is dominated by the control parameter rather than by
the size.
\end{thebibliography}
\end{document}